# Towards a second generation of metascintillators using the Purcell effect

A. Shultzman[a], R. Schütz[a], Y. Kurman[a], N. Lahav[a], G. Dosovitskiy[a], C. Roques-Carmes[b], Y. Bekenstein[a], G. Konstantinou[c], R. Latella[c], L. Zhang[c], F. Loignon-Houle[d], A.J. Gonzalez[d], J.M. Benlloch[d], I.Kaminer[a], and P. Lecoq[c,d,e]

*Abstract—* This study focuses on advancing metascintillators to break the 100 ps barrier and approach the 10 ps target. We exploit nanophotonic features, specifically the Purcell effect, to shape and enhance the scintillation properties of the first-generation metascintillator. We demonstrate that a faster emission is achievable along with a more efficient conversion efficiency. This results in a coincidence time resolution improved by a factor of 1.3, crucial for TOF-PET applications.

*Index Terms—*

## I. INTRODUCTION

THE future generation of radiation detectors is more and more pushing for better timing performance for a wide range of applications, such as particle identification in nuclear physics and high energy physics detectors, high resolution finely segmented hadronic calorimetry, precise event time tagging in high luminosity accelerators, time-of-flight (TOF) techniques for PET (Positron Emission Tomography) cameras and a number of photonic applications based on single photon detection.

In particular, recent trends in PET make use of the TOF information to increase the signal-to-noise ratio (SNR) in the reconstructed image and reduce image artifacts in case of incomplete angular coverage for tomographic reconstruction. For this purpose, a coincidence time resolution (CTR) in the range of 100 ps FWHM or better is desired, corresponding to 1.5 cm along the line of response (LOR) between the two detectors in coincidence [1]. The state-of-the-art in commercially available TOF-PET scanners is now at the level of 200 ps [2]. Despite being a significant improvement over standard PET cameras, this precision does not yet allow a direct 3-D reconstruction of a PET image, which is the ultimate goal and would require a CTR of about 10 ps for a spatial resolution of 1.5 mm along the LOR [3]. Such a precision would allow an on-line image reconstruction with unprecedented SNR, leading to a sensitivity gain of about 20, as compared to the state-of-the-art (Fig. 1). The clinical and economic impact of this sensitivity gain is enormous, as it would allow faster exams, better diagnostic with improved image quality, much lower radioactive dose exposure for the patients, opening the way to a larger deployment of PET scans, in particular in the pediatric, neonatal and even prenatal domains.

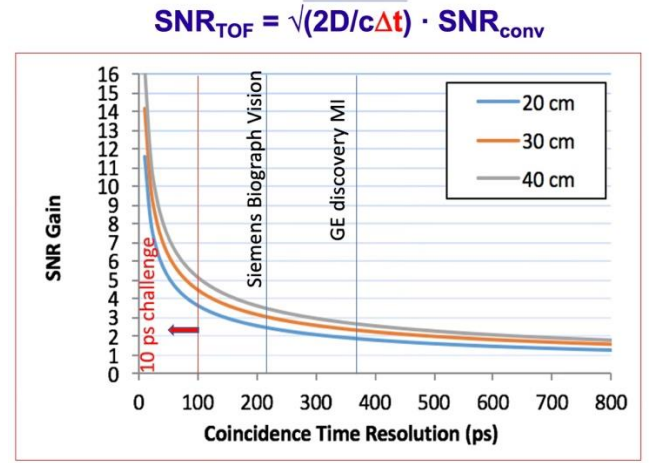

**Fig. 1.** SNR gain versus coincidence time resolution Δt for 3 different diameters of the region of interest. Two commercial PET scanners are shown, the GE discovery MI with a CTR of 375 ps and the Siemens Biograph Vision with a CTR of about 200 ps. The red area below 100 ps is the one investigated with the second generation of metascintillators using the Purcell effect.

## II. THE METASCINTILLATOR CONCEPT

One of the most critical components in TOF-PET instrumentation is the scintillation crystal, which converts the energy of the 511 keV γ-rays into a bunch of optical photons through a complex chain of conversion, transport and energy relaxation mechanisms, which can last several nanoseconds with large statistical fluctuations. In spite of

The metascintillator concept presented in this manuscript was based on research first developed within the ERC Adv project TICAL: 4D total absorption Time Imaging CALorimeter, P. Lecoq (grant agreement no 338953). We also acknowledge the Pazy Foundation and Technion's Helen Diller Quantum Center for their support in developing the nanophotonic concepts.
*Corresponding author: P. Lecoq (paul.Lecoq@cern.ch)*

[a] Technion – Israel Institute of Technology, Haifa, Israel
[b] Stanford, CA, USA
[c] Metacrystal S.A., Geneva, Switzerland
[d] Instituto de Instrumentación para Imagen Molecular (I3M), Centro Mixto CSIC—Universitat Politècnica de València, 46022 Valencia, Spain
[e] CERN, Geneva, Switzerland

many efforts, in particular using co-doping strategies to reduce the delay between the creation of the hot electron-hole pairs and the capture of the resulting slow charge carriers by the luminescent centers after their multiplication and relaxation in the medium [4]-[6], standard scintillation mechanisms in inorganic scintillators are unlikely to produce a scintillation photon rate large enough to break the present barrier of about 200 ps CTR in a realistic PET scanner. Two approaches are presently being investigated to boost the timing resolution of scintillator-based X-ray and γ-ray detectors. The first one consists of exploiting the few Cerenkov photons (< 20, most of them emitted in the UV spectral region with poor detection efficiency) produced by the recoil electron from the photoelectric γ-ray interaction in the medium. The other one is based on the concept of metascintillators introduced in 2008 [7] and first tested in 2017 [8]. It is based on composite scintillator topologies allowing the sampling of the recoil electron produced by the γ-ray conversion in dense scintillator regions in much faster scintillators, such as organic, cross-luminescent or nano-scintillators [9], as shown on Fig. 2.

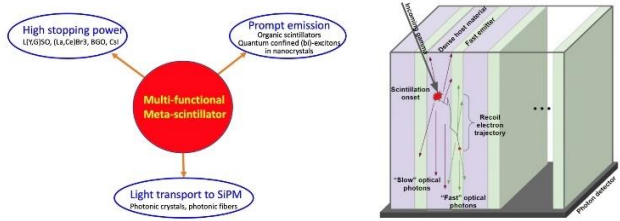

**Fig. 2**. The Metascintillator concept. This figure is adapted from [7,9] with the authors' authorization.

A first generation of metascintillators has been developed, combining in different geometries the high stopping power of BGO or LYSO crystals with the fast emission of plastic scintillators [10] and/or cross-luminescent crystals, such as BaF2 [11]-[13]. In particular, a semi-monolithic metascintillator design (SMMS), allow to determine the depth of interaction (DOI) and further improve the CTR through a time-DOI correction (Fig. 3) [14]. This first generation confirmed the potential of the metascintillator concept to overcome the limitations of standard scintillators, allowing to reach with realistic PET detector dimensions close to 200 ps CTR with BGO-based metacintillators and close to 100 ps CTR with LYSO-based metascintillators, both combined with either plastic EJ232 or BaF2, each time with a subset of events with even much better CTR.

Our ongoing work on a second generation of metascintillators is aiming at breaking the 100 ps barrier and at progressively approaching the ultimate 10 ps target, taking advantage of nanophotonic features allowing to considerably increase the coupling of the electromagnetic wave associated to a particle traversing a medium to the optical states in this medium. Hyperbolic metamaterials have already been proposed to boost the photon conversion efficiency in well-defined regions of SiPM photodetectors, an additional key to improve the timing resolution of scintillators-based ionization radiation detectors [15]. Another possibility is to take advantage of the Purcell effect in polaronic structures engineered in the crystal to create local electromagnetic cavities and increasing the local density of optical states in the material.

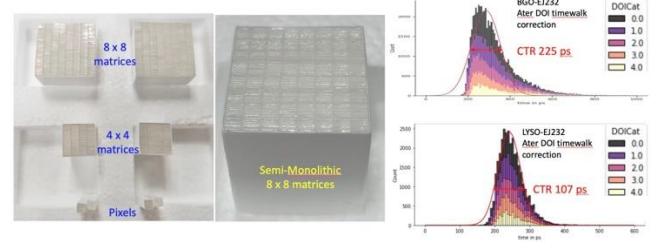

**Fig. 3.** Examples of a few first generation metascintillator configurations and corresponding simulated SMMS CTR results against a LYSO:Ce,Ca reference crystal, with DOI correction. The 5 DOI categories, 3 mm deep each, are defined from 0.0, at the γ-ray entry face of the SMMS to 4.0, close to the SiPM. This figure is adapted from [14] with the authors' authorization.

### III. THE UNDERLYING MECHANISM: PURCELL EFFECT

The Purcell effect describes the enhancement of the spontaneous emission rate of localized point-like emitters, such as atoms, molecules, and quantum dots, due to their electromagnetic surroundings [16,17]. This effect is typically observed when these emitters are positioned within a cavity that has a maximized quality factor (Q) and a minimized light mode volume (V), as the Purcell factor ($F_P$), which represents the emission rate enhancement, is directly proportional to Q/V. This concept was integrated into the field of nanophotonics by Yablonovitch [17].

The Purcell factor can be computed based on the dyadic Green's function, which encompasses the behavior of the whole electromagnetic environment [18]. The spontaneous emission rate enhancement, that is, the Purcell factor of a local dipole, is defined by $F_P = \Gamma/\Gamma_0$, where $\Gamma_0$ is the emission rate of the same dipole in free space.

Using the dyadic Green's function, $\overleftrightarrow{G}$, we can express the partial local density-of-states (LDOS) of a dipole **p** with frequency ω positioned at $\mathbf{r}_0$, by [16]

$$\rho_\mathbf{p}(\mathbf{r}_0, \omega) = \frac{6\omega}{\pi c^2}\left[n_\mathbf{p} \cdot \text{Im}\{\overleftrightarrow{G}(\mathbf{r}_0, \mathbf{r}_0, \omega)\} \cdot n_\mathbf{p}\right] \quad (1)$$

where $n_\mathbf{p}$ is a unit vector in the direction of the dipole, and c is the speed of light in vacuum. One can calculate the total LDOS ρ, by assuming the dipoles are distributed isotropically, and averaging over all dipole orientations. The LDOS is also related to the total power dissipated by the dipole $P_\mathbf{p}$, with

$$P_\mathbf{p} = \frac{\pi\omega^2}{12\epsilon_0}\rho_\mathbf{p}(\mathbf{r}_0, \omega) \quad (2)$$

where $\epsilon_0$ is the vacuum permittivity. Finally, the spontaneous emission rate of a dipole $\Gamma_\mathbf{p}$, is proportional to its LDOS according to

$$\Gamma_\mathbf{p} = \frac{\pi\omega}{3\hbar\epsilon_0}|\mathbf{p}|^2 \rho_\mathbf{p}(\mathbf{r}_0, \omega) \quad (3)$$

where $\hbar$ is Plank's constant. These relations create a link between the quantum and classical formalisms of the dipole, as the enhancement in spontaneous emission rate (the Purcell factor) is equal to the enhancement in the power dissipation

$$\Gamma_\mathbf{p}/\Gamma_{\mathbf{p},0} = P_\mathbf{p}/P_0 \quad (4)$$

where $\Gamma_{\mathbf{p},0}$ and $P_0$ refer to the free space values (which are independent of the dipole's orientation). The link formed in Eq. (4), also removes the explicit dependence on the dipole $\mathbf{p}$.

The enhancement in the spontaneous emission rate, directly enhances the scintillation decay time, through $\tau_d = 1/\Gamma$.

In scintillation, the emission points are dispersed over a volume significantly larger than the wavelength of the radiation, leading to an increased mode volume. Furthermore, lower values of Q are necessary for effective light outcoupling (i.e., the transmission of light from the device to the photodetector through air or an optical coupling material). These conditions contrast sharply with the typical circumstances where the Purcell effect is utilized, which generally require a small cavity volume and a large Q. Thus, although we leverage the same core concept of the Purcell effect, namely the enhancement of spontaneous emission rate with optical environment engineering, we employ a different strategy.

## IV. NANOPHOTONIC SCINTILLATORS

Several recent studies have unveiled the concept of nanophotonic scintillators, demonstrating how the nanophotonic structures can be harnessed to amplify and mold spontaneous emission following excitation by high-energy particles. One pathway that was explored involves improving the light extraction of the scintillator emission using photonic crystals as impedance-matching layers between the scintillator and the photodetector [19–22]. In these works, the actual emission from the scintillator has always been considered an intrinsic property of the material.

A recent work introduced the Purcell scintillators [23] and demonstrated the ability to manipulate the photonic LDOS of a scintillator for the first time, thereby boosting the spontaneous emission rate in scintillation and cathodoluminescence. This work shows the enhanced scintillation efficiency and timing that can be achieved through the innovative design of artificial nanophotonic structures. These structures are crafted from intrinsic scintillators and other dielectric materials and offer a considerable improvement compared to a conventional bulk scintillator. The design of the entire structure incorporates features on length scales that are comparable to the scintillation emission wavelength, which is in the range of hundreds of nanometers. The methodology diverges from the mentioned above traditional applications of nanophotonics in the scintillators context to manipulate pre-existing scintillation emissions. Instead, it amplifies the intrinsic emission process of the scintillator by leveraging the Purcell effect. This technique allows us to enhance the scintillator's emission into directions that can be easily detected, while simultaneously suppressing undetectable emissions.

The contribution of the nanophotonic Purcell scintillator was demonstrated over a vast range of applications, from enhancing the efficiency in light sources [24], to improving the imaging characteristics for detection applications and overcoming fundamental limitations in conventional scintillators [25].

A Purcell nanophotonic scintillator is made of alternating layers of a traditional scintillator and another dielectric material (with different optical properties). Choosing layers' thicknesses on the length scale of the scintillator emission wavelength, will modify the electromagnetic environment of the emitters inside the scintillator layers, and allow for a Purcell enhancement of their spontaneous emission rate. In Fig. 4, we show the dependence of the theoretical Purcell factor on the emitter's orientation, for emitters positioned at different locations inside the multilayer structure. For this example, we consider a multilayer structure containing 5 pairs of the scintillator ($CsPbBr_3$ perovskite nanocrystals) and dielectric (silica). The thicknesses are on the scale of the central wavelength of the emitters' spectrum, which is at 505 nm. We observe that emitters with different locations and orientations, respond differently to the structure. The Purcell factor is computed based on the energy dissipated from the emitters, as demonstrated in Eq. (4).

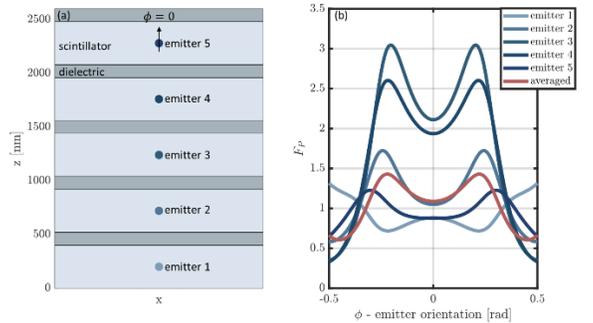

**Fig. 4.** Purcell factor of a single emitter at different locations and dipole orientations inside the multilayer nanophotonic scintillator. (a) Illustration of the multilayer structure containing five pairs of layers. The presented emitters are located at the center of each scintillator layer. For practical applications, the actual distribution of the emitters usually follows an exponential profile according to the absorption coefficients of the materials to the incoming radiation. (b) The Purcell factor of the emitters as a function of their orientation inside the structure. The red curve is the average over the entire volume of the structure.

The angular dependence of the Purcell factor has important implications for many applications, such as imaging applications where this concept can be leveraged in order to direct the emitted light toward the sensor [26]. However, the dependence on the emitters' orientation is not the only phenomenon that contributes to the angular spontaneous emission rate enhancement. There are additional effects caused by the fact that the scintillator consists of an ensemble of emitters distributed over a large volume with a range of frequencies. This average ensemble mixes the contributions of different dipoles, which results in a new angular signature. Moreover, the outcoupling of the field outside the structure needs to be considered using transmission coefficients, which are also angular dependent.

The effective spontaneous emission rate enhancement can be expressed as [27]

$$\Gamma_{\text{eff}}(\theta) = \int dzd\omega\, G(z)\Gamma_o(\theta, \omega)Y(\omega) \quad (5)$$

where $\Gamma_o(\theta, \omega)$ is the outcoupled emission rate that includes the transmission coefficient. In addition, $G(z)$ is the emitter absorption profile in the structure (which depends on the material composition of the layers and on the incoming radiation), and $Y(\omega)$ is the emitters' spectral distribution (normalized to 1).

A first measurement of this phenomenon in the outcoupled emission rate, which considers the averaging overall emitters locations and frequencies, was presented in a recent work [28].

V. USE OF NANOPHOTONIC CONCEPT FOR X-RAY AND $\gamma$-RAY DETECTION APPLICATIONS

*A. Efficiency and resolution enhancement for X-ray imaging*

The potential of Purcell nanophotonic scintillators to augment the imaging capabilities of scintillator-based detectors is an exciting prospect that has been first proposed for X-ray imaging applications [25].

For X-ray imaging applications, the key properties of nanophotonic Purcell scintillators are their efficiency and spatial resolution. The efficiency, $\eta$, is defined as the enhancement in the creation rate of detectable photons, compared to a bulk scintillator having the same dimensions as the Purcell-based metascintillator, computed via

$$\eta = \frac{\int d\theta \sin\theta\, \Gamma_{\text{eff}}(\theta)}{\int d\theta \sin\theta\, \Gamma_0(\theta)} \quad (6)$$

with $\Gamma_0$ being the spontaneous emission rate of the bulk scintillator. The efficiency compares the rate of light creation, averaged over all the modes. To calculate the enhancement in the total number of detectable photons, one needs to integrate the rate over the relevant time period, weighting according to the temporal distribution of the X-ray source. The resolution on the other hand, is related to the ability to distinguish between small features of the imaged object, and is computed based on the system's modulation transfer function (MTF).

The efficiency and resolution of all scintillators and phosphor screens are inherently constrained by the nature of spontaneous emission [26, 27]. Traditional scintillators and phosphors emit light isotropically from the points of X-ray conversion within the material. This leads to further dispersion of light within the material before it reaches photodetectors, thereby reducing the spatial and timing resolutions of the imaging system. Such limitations essentially create a trade-off that is universally applicable to all scintillator-based imaging applications. As depicted in Fig. 5, thicker scintillators are more efficient due to their ability to absorb more radiation and subsequently generate more visible light. However, this comes at the expense of resolution, as light spreads further within the material. Conversely, thinner scintillators can achieve superior resolution, but with reduced efficiency due to their decreased radiation absorption. Using our inverse-designed Purcell nanophotonic scintillators [25], this trade-off can be overcome, by designing a structure achieving high resolution and high efficiency simultaneously. This is obtained by optimizing over the thicknesses of the multilayer nanostructure, to direct the spontaneously emitted light toward the photodetector, which in addition to improving the resolution of the system, also inhibits radiation into unwanted directions, thus increasing the detection efficiency of the fast emission from the Purcell-based nanostructure.

Fig. 5 presents the fundamental trade-off in uniform phosphor screens [25]. We have designed a multilayer nanostructure with alternating layers of P43 phosphor and silica, for enhancing emission centered around 450 nm. P43 phosphor screen is composed of $Gd_2O_2S$:Tb, and is used in a wide range of imaging applications to convert energetic electrons or X-rays into green light. We observe that the multilayer structure is no longer constrained to the trade-off between efficiency and resolution observed in uniform phosphor screens.

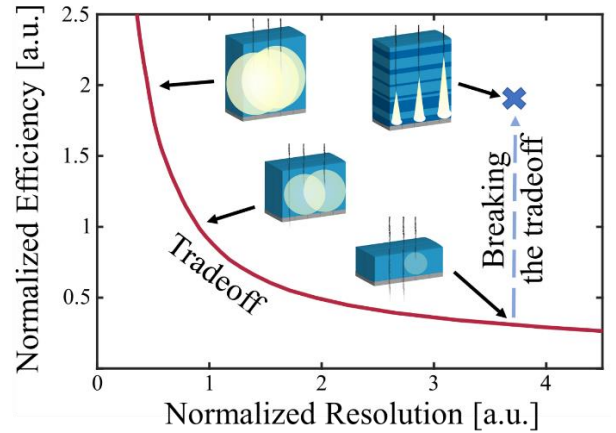

**Fig. 5. Breaking the trade-off between efficiency and resolution in scintillators.** The trade-off between efficiency and resolution in homogeneous phosphor screens (red line). The axes are normalized with respect to the efficiency and resolution of the standard structure. The inverse-designed nanophotonic structure achieves simultaneous high efficiency and high resolution. This figure is reproduced from [25] with the authors' authorization.

## B. Temporal enhancement for γ-ray imaging

Since the launch of the 10 ps Time-of-Flight PET challenge [28] several research groups have published CTR results at the level of 100 ps [29 to 32]. Moreover, the demonstration of a direct non-tomographic image reconstruction on an idealistic setup with a CTR of 30 ps has been made [33, 34]. However, all these tests have been made with LYSO bulk scintillators in perfect laboratory conditions, either with short crystals, or readout conditions not exportable to realistic PET scanners. To surpass the 100 ps CTR barrier and edge closer to the 10 ps milestone, a significant amount of prompt photons (detected in the first 10 ps) are needed, whereas with LYSO, the fastest scintillator used in PET scanners, less than 1 photon can be detected in the first 10 ps. In order to have at least several tens of prompt photons readout in the first 10 ps, allowing to achieve a CTR approaching 10 ps in realistic conditions for a PET scanner, we propose to combine nanophotonic Purcell scintillators within metascintillators. As demonstrated in Fig. 6, the second generation of the metascintillator combines in different geometries the high stopping power of BGO or LYSO crystals with the Purcell-enhanced multilayer scintillators.

We design the nanophotonic scintillator to comply with a few engineering constraints. First, the nanophotonic structure should not hinder the emitted light from the slow layers. Therefore, the layers of the metascintillators are perpendicular to the sensor entrance and photodetector faces. This allows the emitted light in the fast component to outcouple to the slow scintillator (with minimal losses), which guides the light to the photodetector. Secondly, we design the Purcell nanophotonic scintillator to enhance the emission into the directions parallel to the layers.

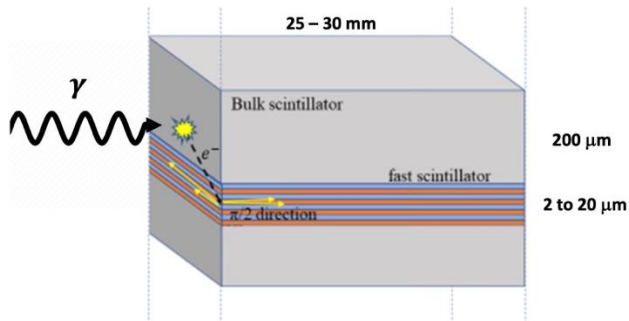

**Fig. 6. Surpassing the 100 ps CTR barrier with the metascintillator concept.** Energy sharing between the bulk BGO or LYSO scintillator crystal layers and a fast photonic crystal made of an optimized multilayer structure exploiting the Purcell scintillator. The desired pixel cross section, for instance to match the SiPM dimensions, will be reached by stacking a number of such multilayer structures.

Purcell scintillators have been proposed as a way to improve the CTR [35]. The CTR resulting from two detectors, scales as $CTR = \sqrt{DTR_1^2 + DTR_2^2}$, where $DTR_{1,2}$ are the detector time resolution, given by [36]

$$DTR \propto \sqrt{\frac{\tau_r \tau_d}{LY}} = \sqrt{\frac{\tau_r}{\Gamma_{eff}}} \quad (7)$$

where LY is the scintillator light yield, $\tau_r$ and $\tau_d$ are the scintillator rise time. And decay time respectively This shows that the DTR, and therefore the CTR, benefit both from the enhancement in efficiency and in effective emission rate. The DTR improvement compared to a uniform scintillator, is of $\sqrt{\Gamma_0/\Gamma_{eff}}$. Using the CTR as our figure of merit for the design of the Purcell nanostructure compared to a bulk scintillator, we aim to minimize $\Gamma_0/\Gamma_{eff}$.

Following the metascintillator concept presented in Fig. 6, we design the Purcell nanophotonic structure that serves as the fast component. The nanostructure is composed of alternating layers of the scintillator ($CsPbBr_3$ perovskite nanocrystal-based composite emitting at a central wavelength of 505 nm), and silica (with refractive index of 1.45). The perovskite nanocrystal has a refractive index of 2.27, while the refractive index of the composite will be smaller depending on the volume load. The thicknesses of the layers are designed to optimize the following objective

$$\max_{\mathbf{d}} \int_{\theta=TIR}^{TIR+\epsilon} d\theta \, \eta(\mathbf{d}) \Gamma_{eff}(\theta; \mathbf{d})/\Gamma_0 \quad (8)$$

where the total internal reflection (TIR) angle is computed based on the layers' refractive indices, $\epsilon$ is an optimization parameter that controls the number of modes we consider reaching the detector, and $\mathbf{d}$ is the vector of thicknesses. The objective is to minimize the DTR, and considers only the relevant emission angles. For the above choice of materials, the TIR angle results in 39.7°.

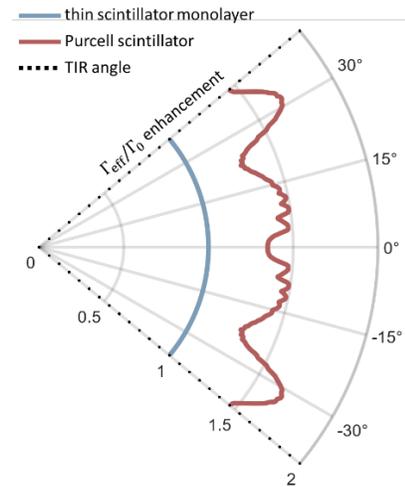

**Fig. 7. Angular spontaneous emission rate enhancement with a Purcell nanophotonic scintillator.** The spontaneous emission rate enhancement of the Purcell

scintillator (red curve) compared to a uniform one (blue curve). The nanophotonic scintillator exhibits faster emission with a maximal enhancement designed to be near the TIR angle, to allow the emitted light to outcouple from the nanophotonic structure and propagate in parallel inside the slow scintillator. This translates to an enhancement in the DTR of $\sqrt{1.7} = 1.3$.

Fig. 7 presents the reduction in spontaneous emission rate enhancement for the excitation of a single γ-ray, $\Gamma_{eff}/\Gamma_0$, for the nanophotonic compared to a uniform scintillator monolayer containing the same amount of scintillator as the nanophotonic structure. The design was done over a multilayer structure with 30 pairs, with the nanophotonic scintillators' optimization tool developed in [25]. The solution of the optimization problem results in a vector of thicknesses for the multilayer nanophotonic scintillator, which is then used to compute the angular dependent DTR. We observe that the emission is specifically enhanced near the TIR angle, allowing the light to be extracted from the nanophotonic structure and propagating in the uniform layers (with lower light absorption). The nanophotonic design reaches an efficiency of $\eta = 1.6$, simultaneously with an improvement of the DTR by a factor of 1.3, as compared to bulk perovskite.

Our choice of perovskite scintillator is motivated by the impressive ongoing efforts to develop this large family of materials for photovoltaic applications, which open the way to scalable cost-effective production methods, as well as to a large flexibility for tuning the scintillation characteristics. As of today, several perovskite crystals have been identified with room temperature scintillation characteristics similar to the EJ232 plastic scintillator we have used in our first generation of metascintillators: about 10 photons / keV and 2ns decay time (see for instance [37, 38]). Moreover, our preliminary measurements indicate that the rise time of these perovskites is much faster than the plastic by at least a factor of 3, which would bring the DTR, and therefore the CTR improvement at the level of $1.3\sqrt{3} = 2.25$ according to formula (7).

Considering the CTR = 107 ps and 120 ps obtained after DOI correction with our first generation of LYSO/EJ232 metascintillators 15 mm and 24 mm long respectively (see Fig. 3 and [14]), replacing the 100 μm thick plastic scintillator by perovskite-based Purcell nanophotonic structures would bring the CTR at the level of 47 ps and 53 ps for 15 mm and 24 mm long metascintillators respectively. Moreover, the much higher density of perovskite, as compared to plastic, would allow to reach the same result with a thinner than 100 μm nonaphotonic scintillator thickness.

Implementing nanophotonic layers in a metascintillator structure allows boosting the number of prompt photons through an increased scintillation efficiency and a directionality of the emission, with a direct impact on the light output and therefore on the timing resolution. On the other hand, we do not expect a significant effect on the energy resolution, because of the stochastic energy sharing between two materials with a priori different light yields. Indeed, the metascintillator concept privileges the γ detection efficiency. The underlying principle is therefore to have the maximum of the incident γ-ray energy deposited in the dense bulk scintillator (typically BGO or LYSO) and to minimize the amount of the generally less dense fast scintillator component to not compromise the overall detection efficiency. The interest of the Purcell-based nanophotonic scintillator is to maximize the light output of the prompt photons, allowing therefore to reach the desired timing resolution with the minimum amount of energy deposited in the nanophotonic material and the smallest impact from the energy sharing correction.

Although the experimental validation of the Purcell enhanced scintillator concept has been established for X-ray imaging [28], fabricating the multilayer nanostructures still remains a challenge. These complexities are particularly pronounced for hard X-rays and γ-rays, which require millimeters to a few centimeters of dense material to be efficiently detected. The integration of nanophotonic scintillator structures in a second generation of metascintillators is the subject of an ambitious collaborative effort involving all the authors of this paper and more results will be reported when the first prototypes will be available.

## VI. CONCLUSION AND OUTLOOK

A first generation of metascintillators, combining in an energy sharing mode the high stopping power of high density and high light yield inorganic scintillators with the fast scintillation from organic, cross-luminescent or nanosized scintillators has shown promises for improving the timing performance in ionization radiation detectors. This technological breakthrough has been documented in several publications in the recent years and reported in this paper, and is particularly important for PET applications, where the timing resolution impacts the effective sensitivity, opening new perspectives in medical healthcare.

The new generation of metascintillators presented in this paper takes advantage of the impressive potential of nanophotonics to engineer the production and transport of optical photons in specifically designed nanostructures. The new concept of nanophotonic scintillators introduced here will replace the fast scintillators layers used in the first generation of metascintillators. The nanoscale multilayer structure exploits the Purcell effect and allows increasing the photonic local-density-of-states and enhancing the spontaneous emission rate in predetermined directions in 4 to 5 times less volume than in the previously nanoscintillator-based scheme. As will be explained in a separate paper in preparation, this opens the way to a paradigm shift in PET imaging, as timing resolutions approaching 10 ps will be within reach, at least for the fraction of events depositing enough energy in the nanophotonic scintillator component of the metascintillators. These events will be used as priors to train the other ones with deep learning techniques on large data sets acquired on phantoms, opening the route to real time molecular imaging without tomographic inversion, as proposed in the 10 ps Time-Of-Flight PET challenge [28].

ACKNOWLEDGMENT

The metascintillator concept presented in this manuscript was based on research first developed within the ERC Adv project TICAL: 4D total absorption Time Imaging CALorimeter, P. Lecoq (grant agreement no 338953). We also acknowledge the Pazy Foundation and Technion's Helen Diller Quantum Center for their support in developing the nanophotonic concepts.